\documentclass[twocolumn,showpacs,preprintnumbers,amsmath,amssymb]{revtex4}

\usepackage{graphicx}
\usepackage{dcolumn}
\usepackage{bm}

\begin{document}

\title{Mixing navigation on networks}
\author{Tao Zhou$^{1,2}$}
\affiliation{%
$^1$Department of Modern Physics, University of Science and
Technology of China, Hefei, 230026, PR China \\
$^2$Department of Physics, University of Fribourg, Chemin du Muse
3, CH-1700 Fribourg, Switzerland
}%

\date{\today}

\begin{abstract}
In this Letter, we proposed a mixing navigation mechanism, which
interpolates between random-walk and shortest-path protocol. The
navigation efficiency can be remarkably enhanced via a few
routers. Some advanced strategies are also designed: For
non-geographical scale-free networks, the targeted strategy with a
tiny fraction of routers can guarantee an efficient navigation
with low and stable delivery time almost independent of network
size. For geographical localized networks, the clustering strategy
can simultaneously increase the efficiency and reduce the
communication cost. The present mixing navigation mechanism is of
significance especially for information organization of wireless
sensor networks and distributed autonomous robotic systems.
\end{abstract}


\maketitle

Information navigation is the fundamental function of all
communication networks. There are two kinds of information: The
nonspecific one, of relevance to broadcasting process, epidemic
spreading and rumor propagation, is desirous to travel all over
the networks, while the specific information focus only on
locating one targeted node. Here we concentrate on the latter
case. In a decentralized file-sharing system, such as GNUTELLA and
FREENET, files are found by forwarding queries to one's neighbors
until arriving the target \cite{Adamic2001}. Without any
navigation, to find a file in those systems is equivalent to a
random-walk search, which is inefficient in large-size networks
\cite{Adamic2001,Noh2004}. The navigation efficiency can be
sharply improved by using some local information, such as the
geographical location of target \cite{Kleinberg2000}, the degrees
of neighboring nodes \cite{Adamic2001,Kim2002}, and \emph{local
betweenness centrality} (LBC) \cite{Thadakamalla2005}. In an
extreme case, if all the nodes know how to deliver the message
along with the shortest path, the highest efficiency can be
achieved with delivery time being equal to the shortest
path-length. However, this ideal navigation system is impractical
in huge-size networks since it requires either a great amount of
external information \cite{Trusina2005} or a huge memory of each
node \cite{Rosvall2003}, which costs too much. Especially, in many
real communication networks, such as Limewire, Kazaa and eDonkey,
the edges can be rapidly rewired \cite{Cholvi2005}. For the
economical and technical reasons, it is hard to design a
navigation system where each node has enough power to detect the
structural or geographical changes, as well as strong
computational ability to dynamically find the shortest paths.

\begin{figure}
\center \scalebox{0.6}[0.6]{\includegraphics{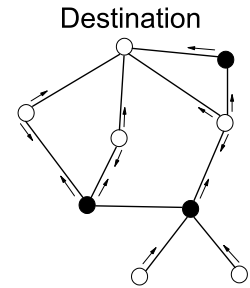}}
\caption{Mixing navigation on networks. The network consists of
two kinds of nodes, forwarders (hollow circles) and routers (solid
circles). As a holder, the forwarder will forward the message to a
randomly selected neighbor, while the router can deliver it one
step towards its destination. The arrows located in each node
represent the possible delivering directions for the given
destination.}
\end{figure}

Many efforts have been made previously on finding highly effective
navigation algorithm with low communication cost. In those
studies, a latent and maybe oversimple assumption is that all the
nodes in the network are functionally equivalent. In this Letter,
we raise a partially centralized navigation system where a tiny
fraction of nodes, called \emph{routers}, know where the shortest
path is, while most nodes, called \emph{forwarders}, can only
randomly forward the message to one of their neighbors. This
mixing navigation mechanism is practicable and necessary in some
significant self-driven systems. Consider a wireless sensor
network \cite{Akyildiz2002,Oqren2004}; its topology varies
dynamically due to the power-exhaustion of some sensors as well as
the changing of frequency channels for security reason. Since the
power of each sensor is limited, it can not be a router especially
for long time. A possible way is that each sensor behaves as a
router for a short period (peer-to-peer way), or, a very few
specific sensors are previously given more power and will be the
routers (partially centralized way). Another typical example is
the distributed autonomous robotic system \cite{Arai2002}, where
each robot moves fast and the direct communication can only be
carried out within a limited horizon like the Vicsek model
\cite{Vicsek1995}. A router must have the ability to detect the
location of target, thus can send the message in the right
direction \cite{Kleinberg2000}. Overmany routers bring high
economic pressure, while without routers the communication will be
inefficient. We expect the embedding of a tiny fraction of routers
could guarantee both the high efficiency and the low cost to the
system. This idea is also inspired by the collective phenomenon of
biological swarm, in which a very few effective leaders can well
organize the whole population \cite{Couzin2005}.

\begin{figure}
\scalebox{0.3}[0.3]{\includegraphics{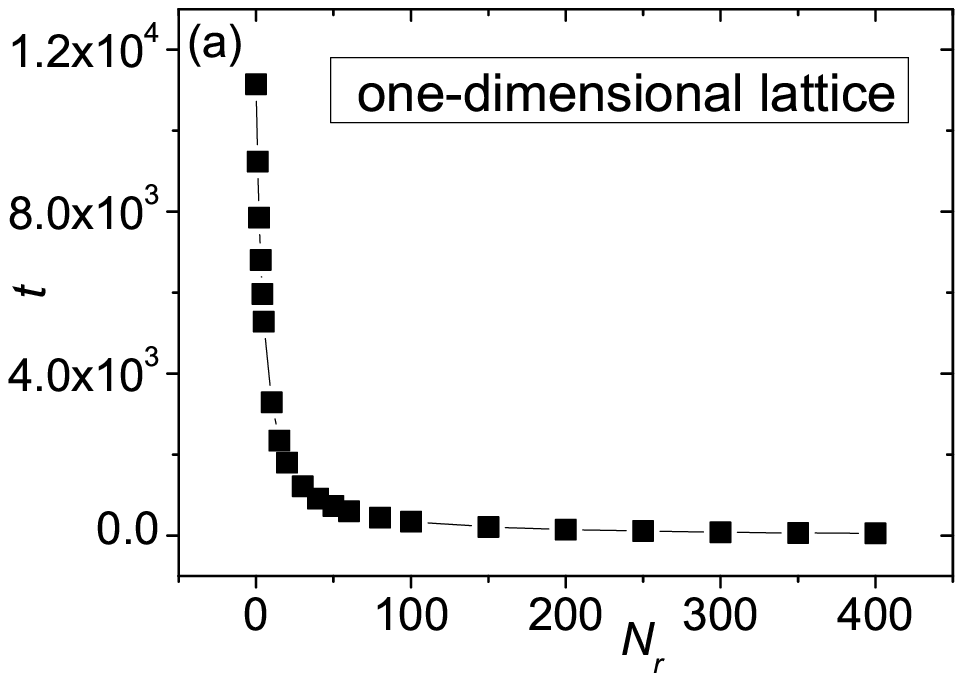}}
\scalebox{0.3}[0.3]{\includegraphics{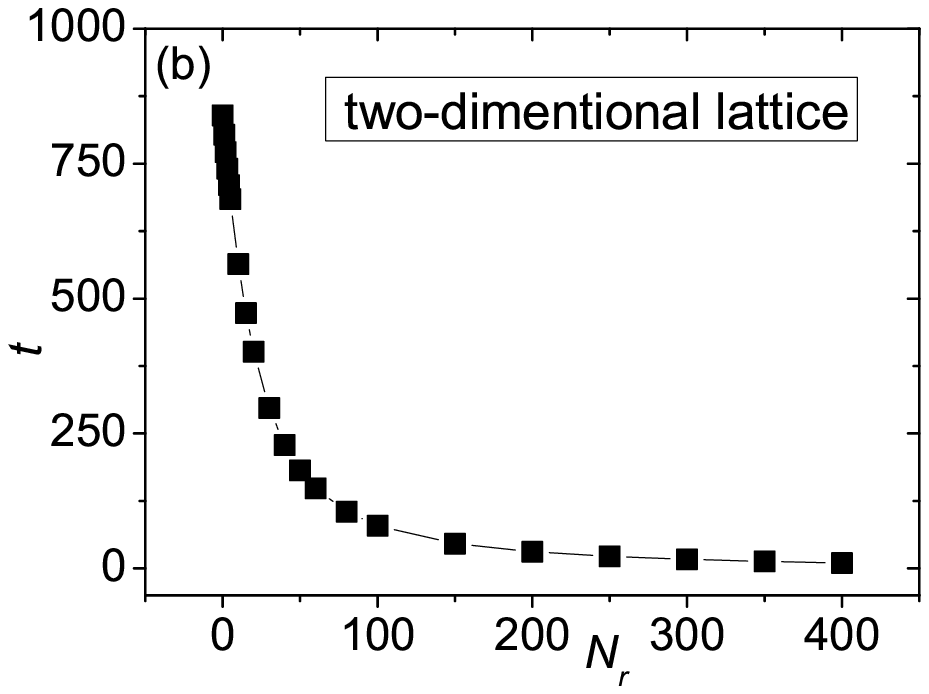}}
\scalebox{0.3}[0.3]{\includegraphics{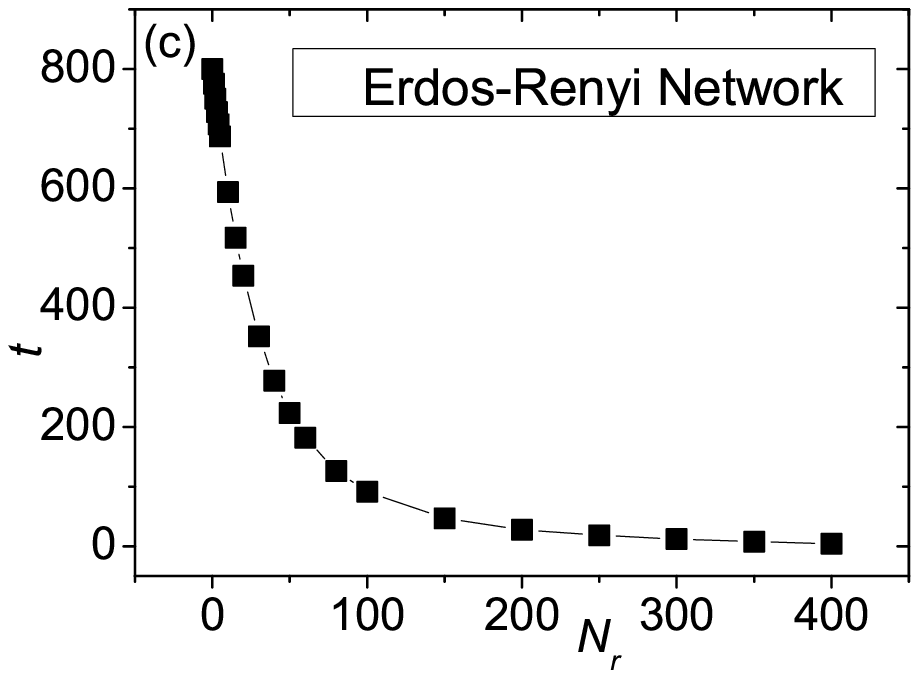}}
\scalebox{0.3}[0.3]{\includegraphics{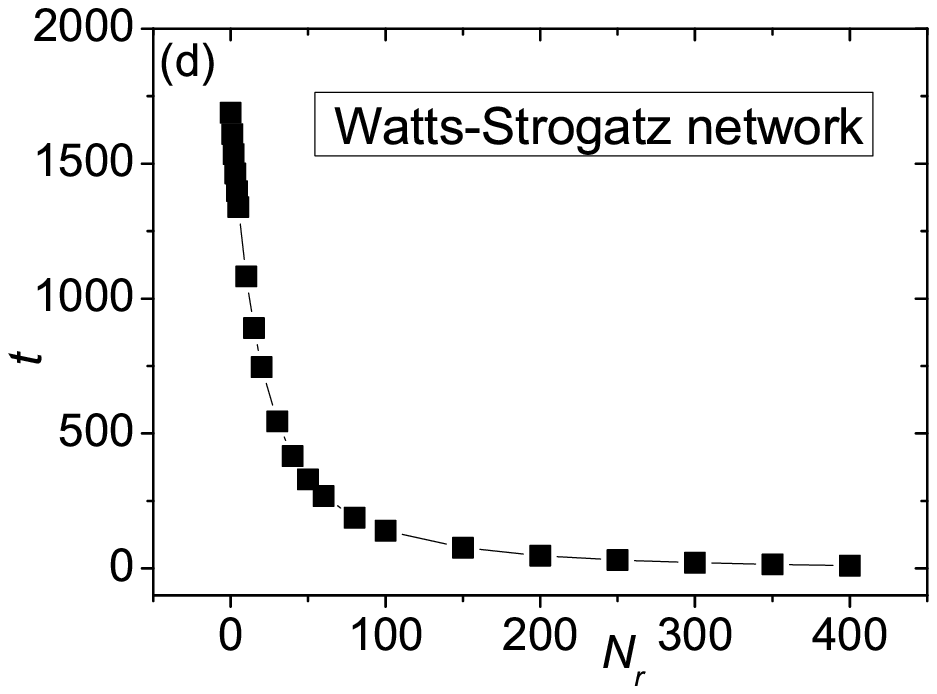}} \caption{Expected
delivery time $t$ vs. $N_r$ with random selection of routers in
(a) one-dimensional lattice, (b) two-dimensional lattice, (c)
Erd\"{o}s-R\'enyi (ER) networks \cite{Erdos1959}, (d)
Watts-Strogatz (WS) networks \cite{Watts1998}. The lattices are of
periodic boundary conditions, and the WS network is generated from
one-dimensional lattice with rewiring probability $p=0.1$. In all
those networks, the size $N=400$ and average degree $\langle k
\rangle=4$ are fixed. All the data points are obtained by
averaging over $10^6$ independent runs.}
\end{figure}

The primary figure of merit for the present model is the expected
delivery time $t$, which represents the expected number of steps
(hop-counts) needed to deliver a message from a random source to a
random destination (see Fig. 1 the rule of mixing navigation).
Taking into account the different measures of cost, we divide the
underlying networks into two classes: One is
\emph{non-geographical networks}, where the Euclidean coordinates
of nodes and the lengths of edges have no physical meaning (e.g.
World-Wide-Web, metabolic networks, etc.). The other is
\emph{geographical networks} having well defined node locations
(e.g. wireless sensor networks, distributed robotic networks,
etc.). In the former case, the cost of router mainly results from
the hardware implementation, since each router needs a large
memory to store the routing table \cite{Yan2006,ex}. Therefore,
the number of routers, $N_r$, is directly used to approximately
measure the cost. In the latter case, usually, the nodes are
moving continuously; since the direct communication is often
bounded with a radius $r_c$, the router has to find out the
location of target, as well as the locations of all its
neighboring nodes with distance $<r_c$. This operation can be
implemented by sending a signal (not message) through a specific
frequency channel to all other nodes and analyzing the feedback,
which requires certain amount of power. Therefore, to save power,
the router may switch its working mode to a simple forwarder
sometimes. The cost, concerning power only \cite{Akyildiz2002},
can be measured by the total time working as a router.

\begin{figure}
\scalebox{0.26}[0.3]{\includegraphics{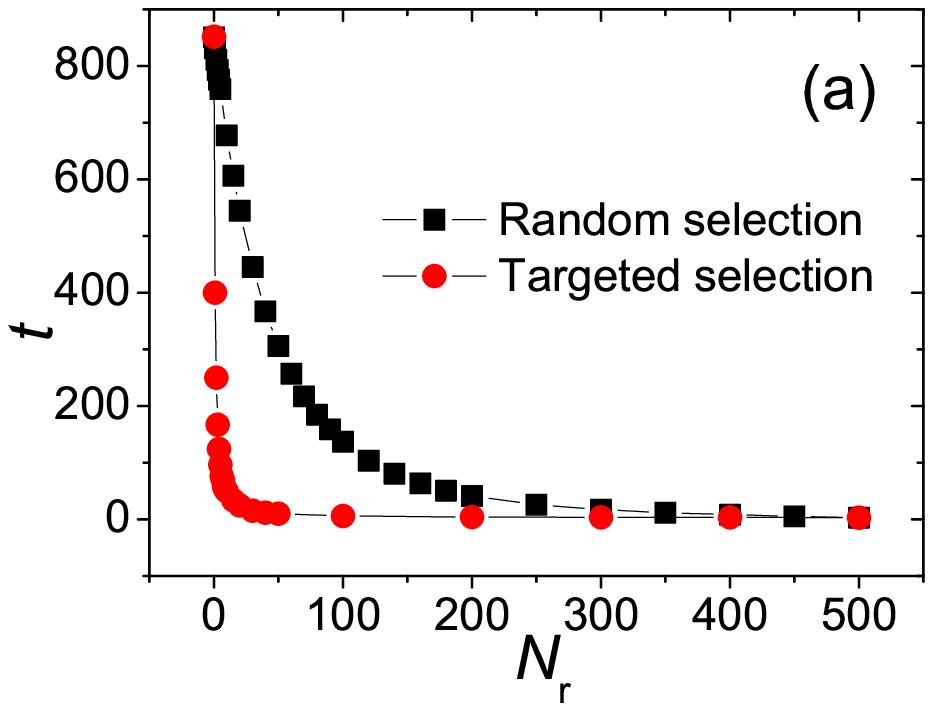}}
\scalebox{0.26}[0.3]{\includegraphics{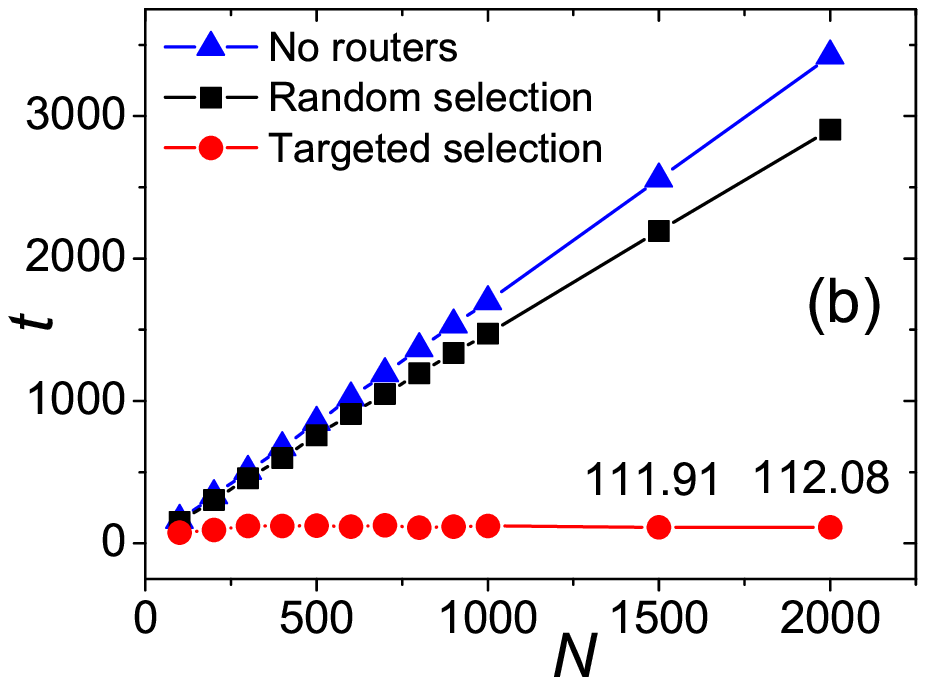}}
\scalebox{0.26}[0.3]{\includegraphics{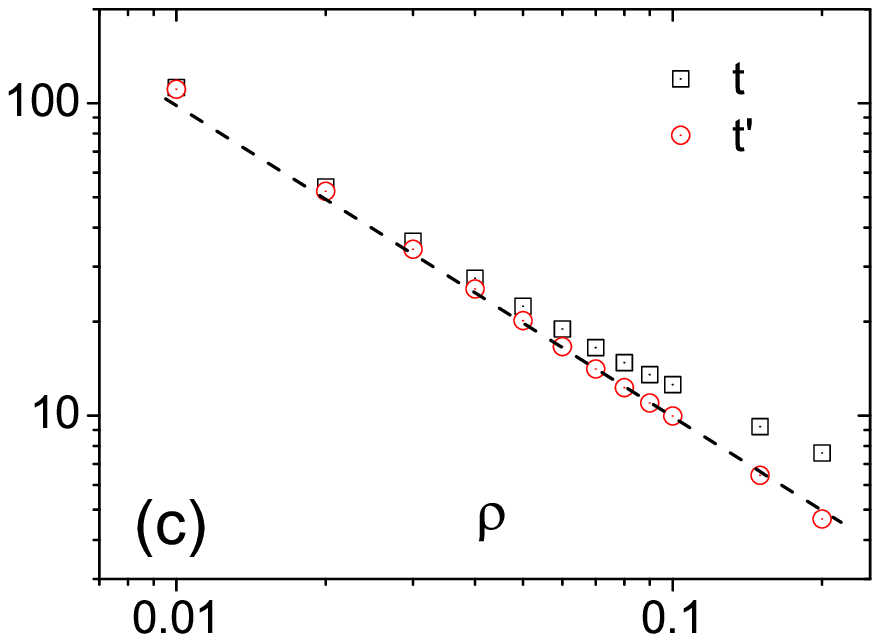}} \caption{(color
online) (a) Expected delivery time $t$ vs. $N_r$ with random
selection (black square) and targeted selection (red circle) on BA
networks with size $N=500$ and average degree $\langle k
\rangle=6$. (b) Scaling behavior of $t$ as the increasing of $N$.
The average degree $\langle k \rangle=6$ and router density
$\rho=N_r/N=0.01$ are fixed. The values of the last two points in
targeted selection is marked. The data points in the two upper
curves can be linearly fitted with slope $1.714\pm0.002$ and
$1.447\pm0.007$, respectively. (c) Expected delivery time $t$
(black square) and rescaled delivery time $t'$ (red circle) as a
function of $\rho$. The straight line is of slope -1. The network
is of $N=2000$ and $\langle k \rangle=6$. All the data points in
(a), (b) and (c) are averaged over $10^7$ independent runs.}
\end{figure}

We start with a trivial method, namely \emph{random selection},
where a few nodes are randomly selected to be routers. Fig. 2
reports the simulation results of some homogenous networks. The
expected delivery time $t$ remarkably decreases after the addition
of a tiny fraction of routers. Then, when $N_r$ gets larger, the
decreasing speed, $-\partial t/\partial N_r$, becomes slower and
the saturation is clearly observed.

\begin{figure}
\scalebox{0.3}[0.3]{\includegraphics{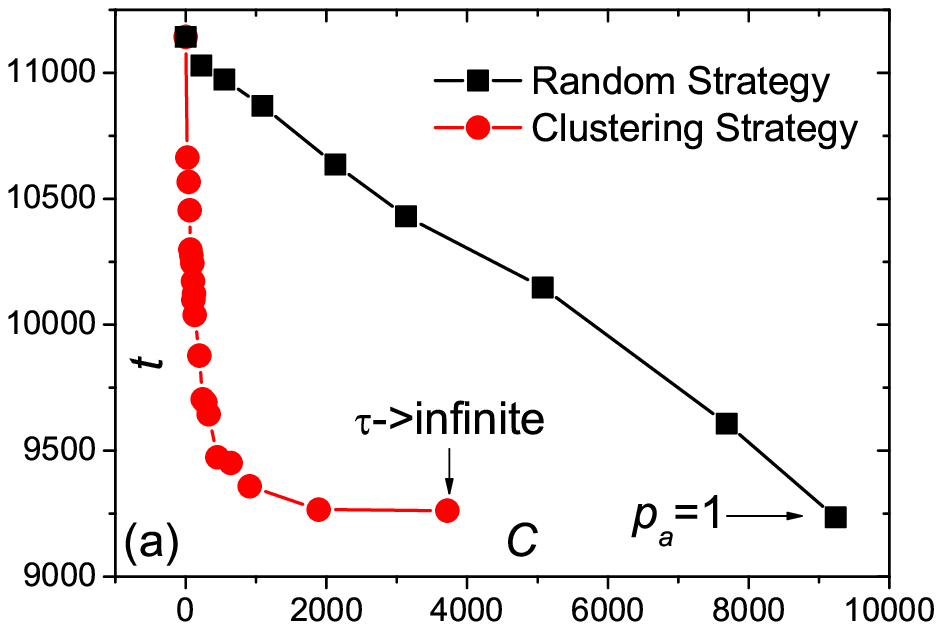}}
\scalebox{0.3}[0.3]{\includegraphics{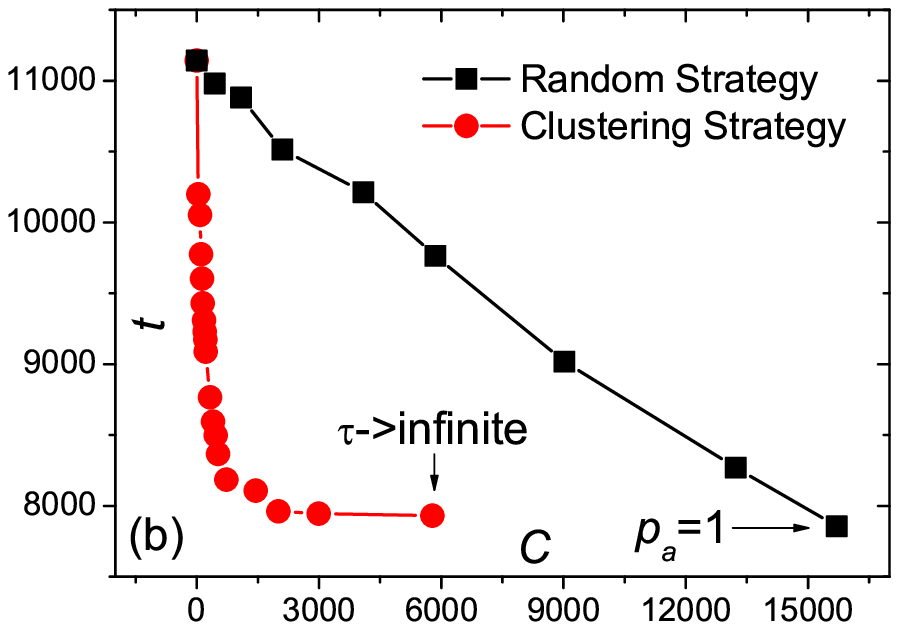}}
\scalebox{0.3}[0.3]{\includegraphics{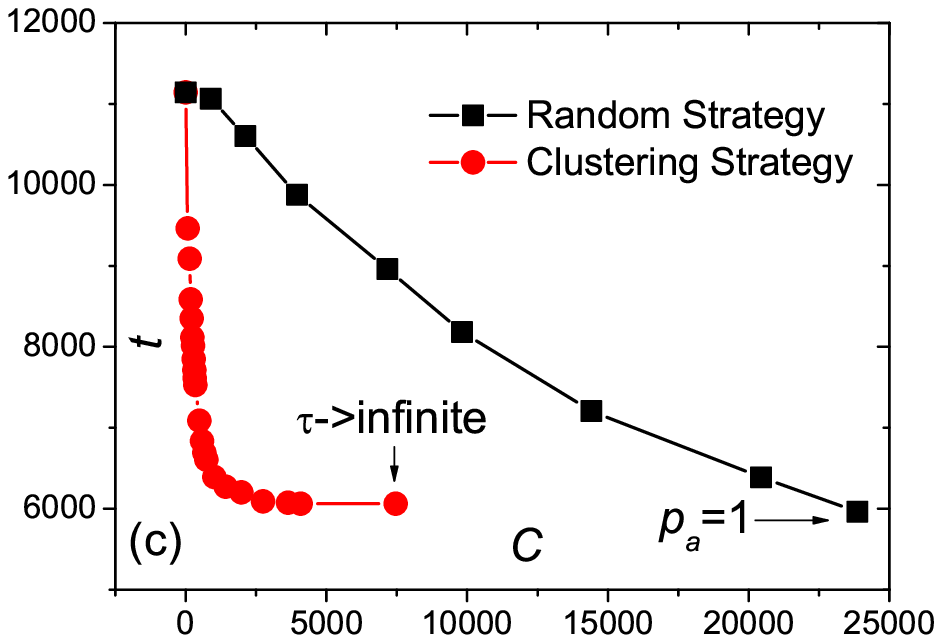}}
\scalebox{0.3}[0.3]{\includegraphics{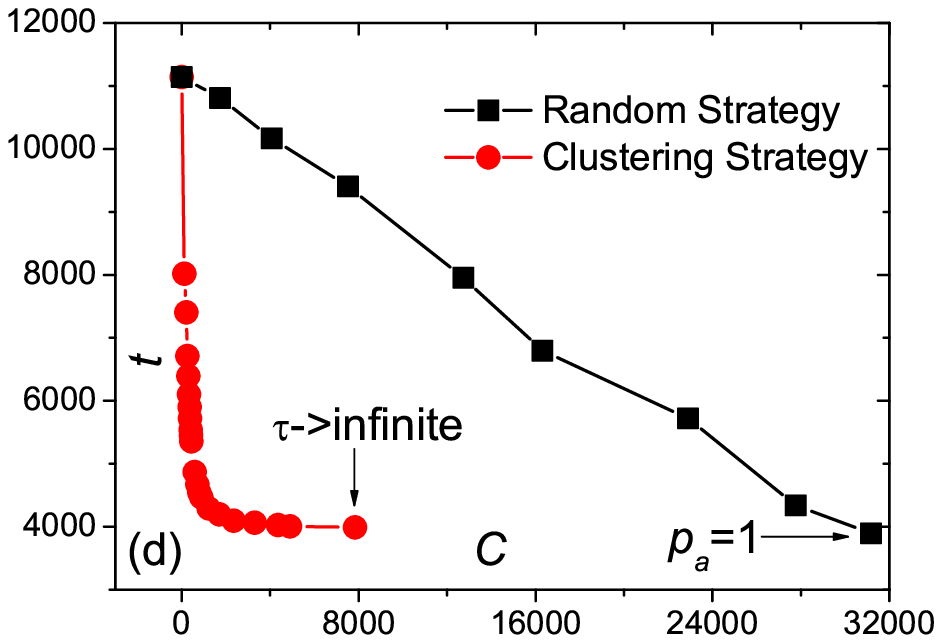}} \caption{(color
online) Expected delivery time $t$ as a function of cost $C$ on
one-dimensional lattice. The number of routers are $N_r=1$ (a), 2
(b), 4(c) and 8 (d), respectively. The $t-C$ relations for random
and clustering strategies are obtained by tuning $\tau$ in
$[1,\infty)$ and $p_a$ in $[0,1]$, respectively. $\tau\rightarrow
\infty$ means the router will never return to inactive state after
it becomes active. $t(\tau\rightarrow \infty)$ is slightly larger
than $t(p_a=1)$ since in the former case when a message visits a
router in the first time, it will be randomly forwarded. The
network size $N=400$ and average degree $\langle k \rangle=4$ are
fixed. All the data points are obtained by averaging over $10^6$
independent runs.}
\end{figure}

Since the majority of real non-geographical networks have
heterogenous degree distribution \cite{Albert2002}, we next
implement this model onto the Barab\'asi-Albert (BA) networks
\cite{Barabasi1999}. Inspired by the prior studies on attack
\cite{Albert2000,Cohen2001} and immunization
\cite{Pastor-Satorras2002}, we propose a \emph{targeted selection}
strategy where $N_r$ nodes with the highest degree are selected to
be routers. As shown in Fig. 3a, compared with the random
selection strategy, the targeted one has much higher efficiency.
With one router added, the delivery time $t$ drops to its half,
and the efficiency can be enhanced about 10 times via only 5
routers. Fig. 3b shows the delivery time as a function of network
size. Without any routers, $t$ scales linearly with $N$, and a
small fraction of randomly added routers will not change its
scaling behavior, but only reduce the growing rate $\partial
t/\partial N$. Surprisingly, under the targeted strategy, even a
very tiny fraction (e.g. $\rho=0.01$) of routers can guarantee a
highly efficient navigation with $t$ almost stable as the
increasing of system size. The scaling behavior can be
analytically predicted in the large-size limit $N\rightarrow
\infty$. If $\rho=0$, the current navigation algorithm degenerates
to random walk with $t\sim N$ \cite{ex2}, and for any $\rho$
larger than the percolation threshold \cite{Callaway2000}, the
network is decomposed into many interconnected forwarder-cores
bounded with routers, and the delivery time consists of two parts:
One accounts for the time randomly walking inside the cores, the
other for the time travelling along with the boundary from the
core containing the source to that containing the destination.
Approximately, the former scales as $t_1\sim 1/\rho$, while the
latter is approximated to $t_2\approx \langle
l\rangle(1+\texttt{log}^\rho_N)$, where $\langle l\rangle$ denotes
the average shortest path-length. When $\rho$ is very small (close
to $1/N$), the contribution of $t_2$ is neglectable even for huge
(but not really infinite) $N$ , thus $t\approx t_1\sim 1/\rho$. As
shown in Fig. 3c, in the log-log plot, $t(\rho)$ can be fitted by
a straight line with slope -1 for very small $\rho$. However, when
$\rho$ gets larger, the departure from $\rho^{-1}$ scaling becomes
visible. To move out the contribution of $t_2$, we use a rescaled
delivery time $t'=t-\langle l\rangle(1+\texttt{log}^\rho_N)$,
which can be well fitted by a straight line with slope $-1.002\pm
0.005$ in the interval $\rho<0.3$. When $\rho$ goes close to 1,
$t_2$ will be dominant thus $t \sim \texttt{ln}N$, as what we
expect in the shortest-path searching algorithm. The details of
analyses will be published in an extending paper.

\begin{figure}
\center \scalebox{0.45}[0.45]{\includegraphics{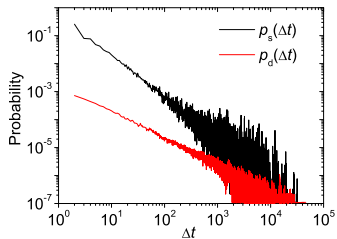}}
\caption{(color online) The time-correlated hitting probability
$p_s$ and $p_d$ as a function of time interval $\Delta t$. The
simulation results are averaged over $10^7$ independent runs, for
a one-dimensional lattice with $N=400$, $N_r=2$ and $\langle
k\rangle$=4.}
\end{figure}

\begin{figure}
\scalebox{0.3}[0.3]{\includegraphics{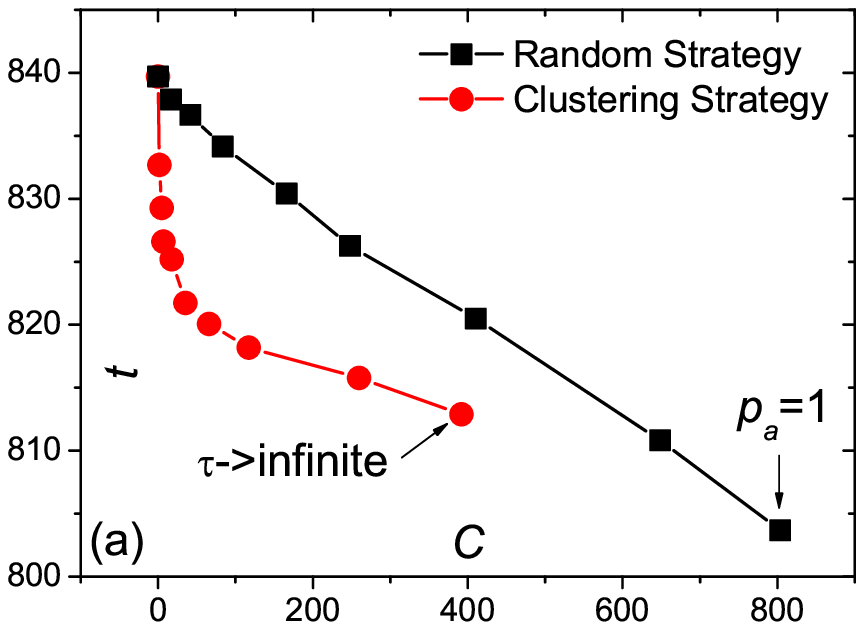}}
\scalebox{0.3}[0.3]{\includegraphics{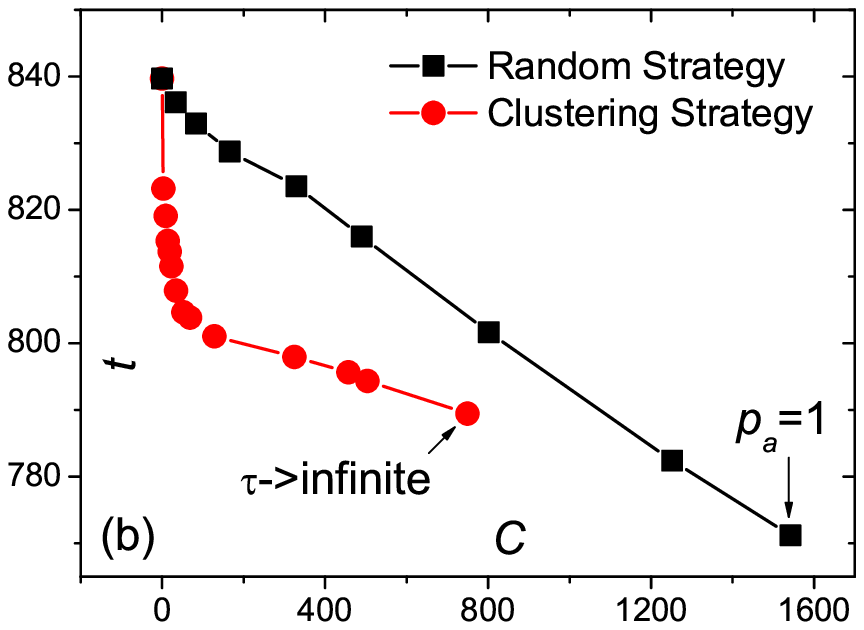}}
\scalebox{0.3}[0.3]{\includegraphics{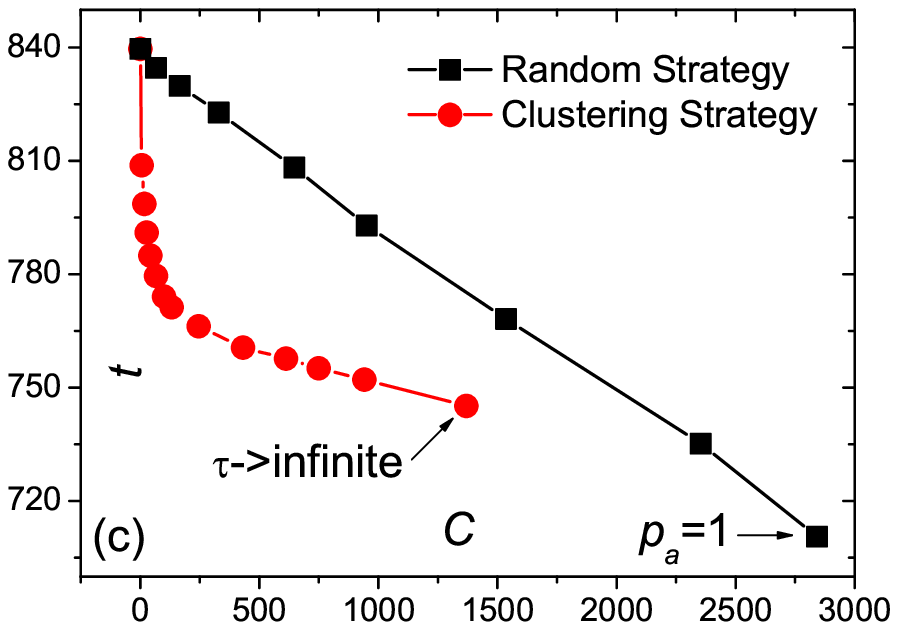}}
\scalebox{0.3}[0.3]{\includegraphics{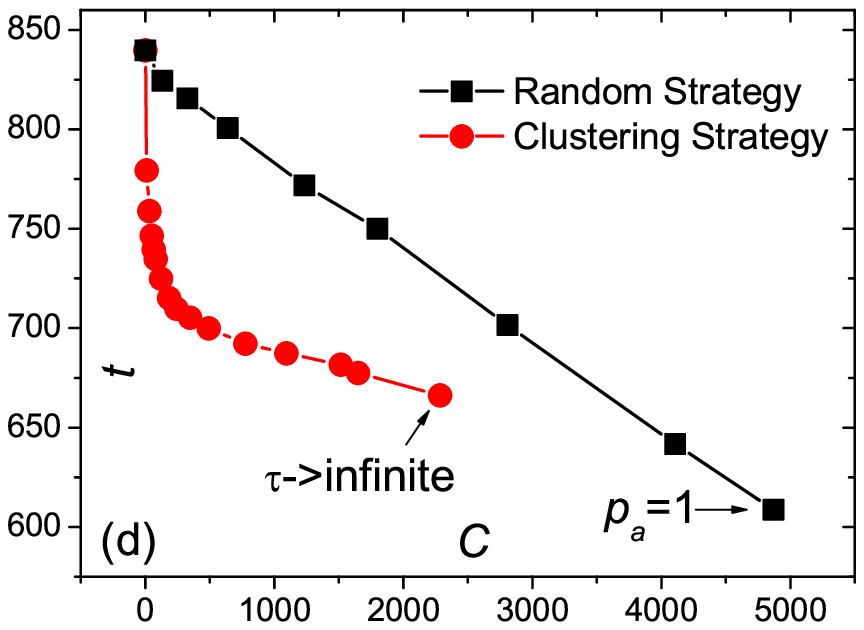}} \caption{(color
online) $t$ vs. $C$ on two-dimensional lattice. The setting of
$N$, $N_r$ and $\langle k\rangle$ are the same as those of Fig.
4.}
\end{figure}

We next explore the navigation on geographical networks. For
simplicity, the network is embedded in a one-dimensional lattice
with periodic boundary condition, and the average degree, which
reflects the horizon $r_c$, is fixed as $\langle k\rangle=4$. The
routers are randomly distributed in the network, each of which can
be in one of the two states: \emph{active} or \emph{inactive}. In
the former state, the router will continuously send/recieve
specific signals to detect the locations of all other nodes thus
can delivery the message one step towards its destination. In the
latter state, the router behaves like a simple forwarder. The
cost, denoted by $C$, is measures by the total active time summing
over all the routers. The transmit time from one node to its
neighbor is assumed to be the same for any message, and counted as
the system time unit. The simplest strategy is to switch randomly,
that is to say, at each time step, each router will be active with
probability $p_a$, where $p_a$ is a constant independent of time.
Given network structure and the number of routers, both the
delivery time $t$ and the cost $C$ are statistically determined by
$p_a$, thus by tuning $p_a$, a curve in $t-C$ (efficiency-cost)
plane can be obtained. Moreover, we propose a novel switching
method, namely \emph{clustering strategy}. In this strategy, if an
inactive router receives a message at time step $T$, it will
forward it to a random neighbor and then becomes active from time
$T+1$ to $T+\tau$. For an active router, it will send this message
one step along with the shortest path to the destination, and keep
active from time $T+1$ to $T+\tau$. If this message will not come
again before $T+\tau$, the router switches to inactive. Initially,
all the routers are inactive. Analogously, by adjusting $\tau$, a
curve in the efficiency-cost plane can be obtained.

\begin{figure}
\scalebox{0.26}[0.3]{\includegraphics{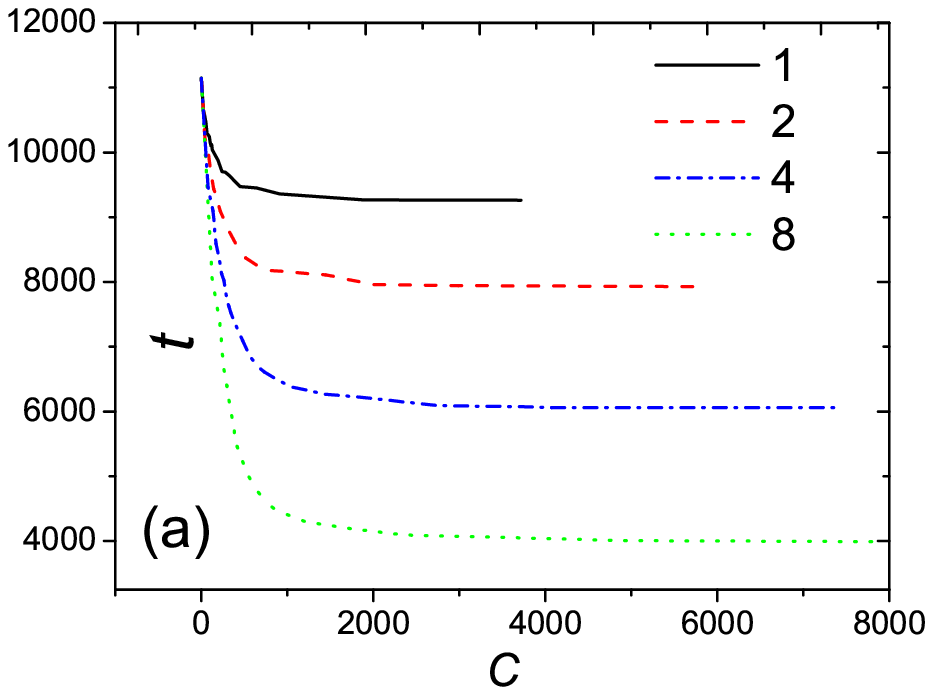}}
\scalebox{0.26}[0.3]{\includegraphics{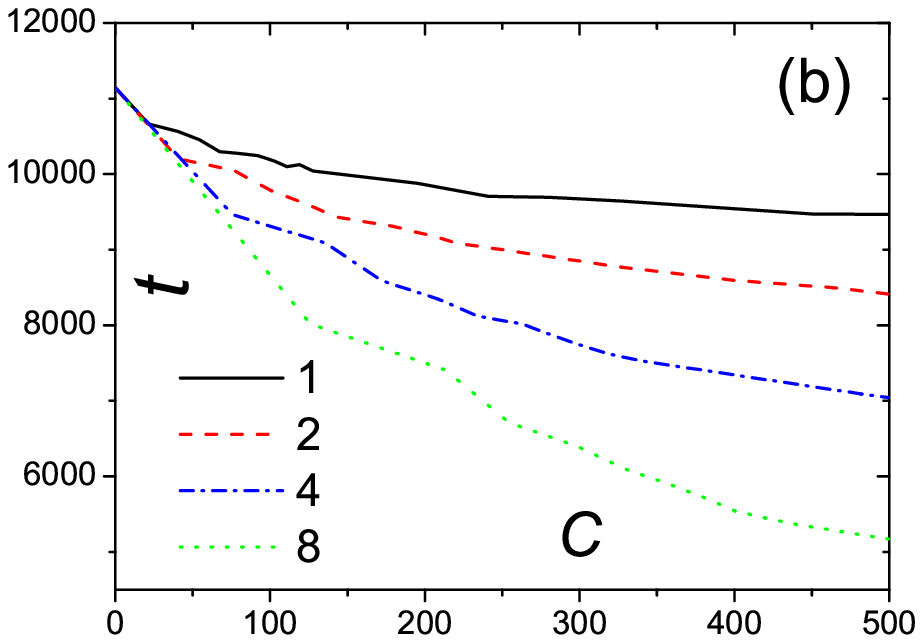}}
\scalebox{0.26}[0.3]{\includegraphics{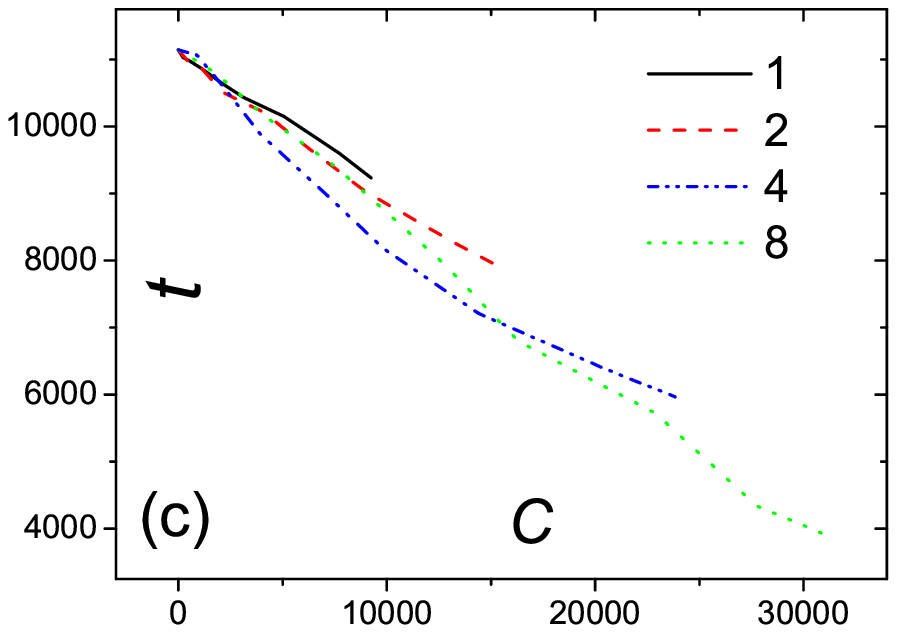}} \caption{(color
online) The relations between $t$ and $C$ for $N_r=1$ (black
solid), 2 (red dash), 4 (blue dash-dot) and 8 (green dot) under
clustering strategy (a) and random strategy (c), respectively. The
panel (b) illustrates the part of (a) for small $C$. The data
shown here are obtained from the same simulation environment as
that of Fig. 4.}
\end{figure}

The simulation results for random and clustering strategies are
shown in Fig. 4. Clearly, by raising the cost (i.e., increasing
$p_a$ and $\tau$, respectively), the navigation efficiency can be
enhanced. With the same cost, the clustering strategy performs
much better than random strategy. It is because the track of
message has a localized effect (also called the phenomenon of
\emph{information clustering}). Generally, after visiting a router
$i$, a message will walk within $i's$ surrounding area during a
certain time period, with much higher probability hitting $i$
again than other ``far away" routers. To measure this localized
effect, we introduce a time-correlated hitting probability
$p(\Delta t)$, which is defined as the probability that the time
interval of two sequent hits is $\Delta t$, where a hit means the
message arriving at a router. Divide $p(\Delta t)$ into two parts
$p(\Delta t)=p_s(\Delta t)+p_d(\Delta t)$, where $p_s(\Delta t)$
($p_d(\Delta t)$) denotes the case if two sequent hits are on the
same router (different routers). Fig. 5 reports the simulation
results of $p_s$ and $p_d$ in one-dimensional lattice. Clearly,
for small $\Delta t$, $p_s$ is $>2$ orders of magnitude larger
than $p_d$, indicating the strongly localized effect. Therefore,
the clustering strategy simultaneously has two advantages:
Firstly, it increases the probability that a router is active when
being revisited, thus can enhance the efficiency; on the other
hand, it avoids the useless activities of some ``far away"
routers, thus can hold down the cost. Fig. 6 reports the
simulation results about $t-C$ relations on two-dimensional
lattice with periodic boundary condition, which also demonstrate
the visible advantage of clustering strategy. However, the
improvement from random to clustering strategy in the
two-dimensional lattice is smaller than that in one-dimensional
case. Actually, the advantage of clustering strategy is more
remarkable in more localized networks. In geographical lattice,
the one having larger scale $N^{1/d}$ and smaller horizon $r_c$ is
more localized, where $d$ denotes the dimension.

Clearly, better efficiency can be achieved by adding more routers.
However, it may also increase the cost. As shown in Fig. 7c, for
the case of random strategy, the four curves for different $N_r$
have almost the same decaying rate. Therefore, if using the
decrement of $t$ resulting from unit cost to judge the strategy,
the addition of routers can not enhance the performance of random
strategy. This finding is hackneyed in many real situation: Given
a strategy, if one wants to gain more, one has to pay more.
Interestingly, it is found that the performance of clustering
strategy can be enhanced by adding more routers. As shown in Fig.
7a and Fig. 7b, the decaying rate of larger-$N_r$ curve is
remarkably higher than that of smaller-$N_r$ curve, indicating
that the clustering strategy with larger $N_r$ can bring more
improvement via unit cost. Although only the cases $N_r=1,2,4,8$
are plotted, this conclusion is valid for all $N_r=1,2,\cdots,N$.
For example, in the extreme case $N_r=N=400$, to reduce delivery
time to $t\approx 150$ only costs $C\approx 375$.

In conclusion, the efficient of mixing navigation in
non-geographical networks is strongly related to the percolation
problem. When $p$ exceeds the percolation threshold, the
underlying network will be decomposed into many small-size
forwarder-cores, guaranteeing the short delivery time. It is the
reason why the targeted strategy can give rise to a highly
efficient navigation with very low communication cost. For
geographical networks, taking into account the information
localization, we proposed a clustering strategy, whose advantage
is more remarkable in more localized networks. The strength of
localization can be measured by the ratio $p_s/p_d$, the higher
ratio indicates the stronger localized effect. Since the hardware
cost of single sensor drops exponentially and power supply becomes
the bottleneck of well communication in huge-size wireless sensor
networks, the clustering strategy, especially the extreme case
$N_r=N$, is of significant importance in practice.

This work is support by the NNSFC under Grant Nos. 10635040 and
10472116.


\begin{thebibliography} {1}
\bibitem{ex2} For a scale-free network with degree distribution $p(k)\sim
k^{-\gamma}$, the expected delivery time of a random-walk
navigation scales as $t\sim N^{3-6/\gamma}$, since for BA
networks, $\gamma=3$, $t$ sacles linearly with the network size
$N$.
\end{thebibliography}
\end{document}